\documentclass[a4paper,12pt]{article}
\usepackage[pctex32]{graphics}
\usepackage{amssymb,amsmath}

\begin{document}
\topmargin 0pt
\oddsidemargin 0mm
\newcommand{\be}{\begin{equation}}
\newcommand{\ee}{\end{equation}}
\newcommand{\ba}{\begin{eqnarray}}
\newcommand{\ea}{\end{eqnarray}}
\newcommand{\fr}{\frac}
\newcommand{\nn}{\nonumber}
\renewcommand{\thefootnote}{\fnsymbol{footnote}}

\begin{titlepage}

\vspace{5mm}
\begin{center}
{\Large \bf Phase transitions of the BTZ black hole in new massive
gravity}

\vskip .6cm
 \centerline{\large
 Yun Soo Myung$^{a}$}

\vskip .6cm

{Institute of Basic Science and Department of Computer Simulation,
\\Inje University, Gimhae 621-749, Korea \\}
\end{center}

\begin{center}

\underline{Abstract}
\end{center}
We investigate  thermodynamics of the BTZ black hole in new massive
gravity explicitly. For $m^2\ell^2>1/2$ with $m^2$ the mass
parameter of fourth-order terms and $\ell^2$ AdS$_3$ curvature
radius, the Hawking-Page phase transition occurs between the BTZ
black hole and AdS (thermal) soliton. For $m^2\ell^2<1/2$, however,
this transition unlikely occurs but a phase transition between the
BTZ black hole and the massless BTZ black hole is possible to occur.
We may call the latter as the inverse Hawking-Page phase transition
and this transition is favored  in the new massive gravity.
\vspace{5mm}


\vskip 0.8cm

\vspace{15pt} \baselineskip=18pt \noindent $^a$ysmyung@inje.ac.kr

\thispagestyle{empty}
\end{titlepage}
\renewcommand{\thefootnote}{\arabic{footnote}}
\setcounter{footnote}{0}
\newpage
\section{Introduction}

 A black hole could be rendered
thermodynamically stable by placing it in four-dimensional anti-de
Sitter (AdS$_4$) spacetimes because AdS$_4$ spacetimes play the role
of a confining box.   Then, it is a natural question to ask  how a
stable black hole with positive heat capacity could emerge from
thermal radiation through a phase transition. This was known to be
the Hawking-Page (HP) phase transition between thermal radiation
(TR) and Schwarzschild-AdS$_4$  black hole (SAdS)~\cite{HP,BCM}. It
has shown one typical example of the first-order phase transition
(TAdS$\to$ small SAdS $\to$large SAdS) in the gravitational system.
In the last two decades, its higher dimensional extension and its
holographic dual to confinement-deconfinement  transition were the
hottest issues~\cite{Witt}.

In order to study the HP phase transition in Einstein gravity, we
need  to know the Arnowitt-Deser-Misner (ADM)
mass~\cite{Arnowitt:1962hi}, the Hawking temperature, and the
Bekenstein-Hawking (BH) entropy. These are combined to give the
on-shell  free energy in canonical ensemble which determines the
global thermodynamic stability. The other important quantity is the
heat capacity which determines the local thermodynamic stability.
Employing the Euclidean action formalism, one easily  finds these
quantities~\cite{dbranes}.  However, a complete computation of the
thermodynamic quantities was limited in fourth-order gravity because
one has encountered some difficulty to compute their conserved
quantities in asymptotically AdS spacetimes.

In three dimensions, either third-order gravity (topologically
massive gravity~\cite{Deser:1981wh}) or the fourth-order gravity
(new massive gravity~\cite{Bergshoeff:2009hq}) is essential to
describe a  spin-2 graviton because the Einstein gravity is a gauge
theory without propagating degrees of freedom. Recently, there was a
significant progress on computation of mass and thermodynamic
quantities by using the Abbot-Deser-Tekin (ADT)
method~\cite{Abbott:1981ff,Deser:2002jk,Kim:2013zha}. One has to
recognize that all ADT thermodynamic quantities except the Hawking
temperature depend on a mass parameter $m^2$. Hence,  for
$m^2\ell^2>1/2$, all thermodynamic properties are dominantly
determined by Einstein gravity, while for $m^2\ell^2<1/2$, all
thermodynamic properties are dominantly  determined by purely
fourth-order curvature term. More recently, it was shown that the HP
phase transition (thermal soliton$\to$ BTZ black hole) occurs for
$m^2\ell^2>1/2$ in new massive gravity by computing off-shell free
energies of black hole and soliton~\cite{Zhang:2015wna}. However,
the role of the massless BTZ black hole was missed.  The former can
be  completely understood if the massless BTZ black hole is
introduced as a mediator.  Furthermore,  the present is a turnaround
time to explore the $m^2\ell^2<1$ case of black hole thermodynamics
if one wishes to study the black hole thermodynamics by employing
the massive gravity theory.

 On the other hand, we would like to mention that the
stability condition of the BTZ black hole in the new massive gravity
turned out to be  $m^2\ell^2>1/2$ regardless of the horizon size
$r_+$, while the instability  condition is given by $ m^2\ell^2
<1/2$~\cite{Moon:2013lea}.  For  $ m^2\ell^2 <1/2$, the BTZ black
hole is thermodynamically unstable because of $C_{\rm ADT}<0$ and
$F^{on}_{\rm ADT}>0$ as well as it is classically unstable against
the metric perturbations. The latter indicates  a perturbative
instability of the BTZ black hole arisen from  the massiveness of
graviton.  It implies  a deep connection between thermodynamic
instability and classical instability for the BTZ black hole only
for the new massive gravity~\cite{Myung:2013uka}. Also, it suggests
that the phase transition for $m^2\ell<1/2$ is quite different from
that of the $m^2\ell>1/2$ case. Here, we wish to explore the
presumed phase transition and it will be compared with the
Hawking-Page phase transition for the $m^2\ell>1/2$ case.

\section{Thermodynamics of the BTZ black hole}

We introduce the new massive gravity (NMG) composed of the
Einstein-Hilbert action with a cosmological constant $\lambda$ and
fourth-order curvature terms~\cite{Bergshoeff:2009hq}
\begin{eqnarray}
\label{NMGAct}
 S_{\rm NMG} &=&S_{\rm EH}+S_{\rm FOT },  \\
 \label{NEH} S_{\rm EH}&=& \frac{1}{16\pi G} \int d^3x \sqrt{-g}~
  (R-2\lambda) \\
\label{NFO} S_{\rm FOT }&=&-\frac{1}{16\pi Gm^2} \int d^3x
            \sqrt{-g}~\Big(R_{\mu\nu}R^{\mu\nu}-\frac{3}{8}R^2\Big),
\end{eqnarray}
where $G$ is a three-dimensional Newton constant and $m^2$ a mass
parameter with mass dimension 2. In the limit of $m^2 \to \infty$,
$S_{\rm NMG}$ recovers  the Einstein gravity, while $S_{\rm NMG}$
reduces to purely fourth-order gravity in the limit of $m^2 \to 0$.
The Einstein equation is given by \be \label{eqn}
R_{\mu\nu}-\frac{1}{2}g_{\mu\nu}R+\lambda
g_{\mu\nu}-\frac{1}{2m^2}K_{\mu\nu}=0,\ee where
\begin{eqnarray}
  K_{\mu\nu}&=&2\square R_{\mu\nu}-\frac{1}{2}\nabla_\mu \nabla_\nu R-\frac{\square{}R}{2}g_{\mu\nu}
        +4R_{\mu\rho\nu\sigma}R^{\rho\sigma} -\frac{3R}{2}R_{\mu\nu}-R^2_{\rho\sigma}g_{\mu\nu}
         +\frac{3R^2}{8}g_{\mu\nu}.
\end{eqnarray}
The  BTZ  black hole solution to Eq.(\ref{eqn}) is given
by~\cite{BTZ-1,BTZ-2}
 \be
\label{2dmetric}
  ds^2_{\rm BTZ}=\bar{g}_{\mu\nu}dx^\mu dx^\nu=-f(r)dt^2
   +\frac{dr^2}{f(r)}+r^2d\phi^2,~~f(r)=-M+\frac{r^2}{\ell^2}
\end{equation}
when satisfying a condition of $1/\ell^2+\lambda+1/(4m^2\ell^4)=0$
with $\ell^2$ the curvature radius of AdS$_3$ spacetimes. Here $M$
is related to  the ADM mass of black hole. The horizon radius $r_+$
is determined by the condition of $f(r_+)=0$.

 On the other hand, the linearized equation to
(\ref{eqn}) upon choosing the transverse-traceless  gauge of
$\bar{\nabla}^\mu h_{\mu\nu}=0$ and $h^\mu~_\mu=0$ leads to the
fourth-order linearized equation for the metric perturbation
$h_{\mu\nu}$
\begin{equation}
\Big(\bar{\nabla}^2-2\Lambda\Big) \Big[\bar{\nabla}^2-2\Lambda
-{\cal M}^2(m^2)\Big] h_{\mu\nu} =0,~~\Lambda=-\frac{1}{\ell^2}
\label{linh}
\end{equation}
which might imply  the two second-order linearized equations
\begin{eqnarray}\label{nmgmeq1}
&&\Big(\bar{\nabla}^2-2\Lambda\Big)h_{\mu\nu}=0,
\\ \label{nmgmeq2}
&&\Big[\bar{\nabla}^2-2\Lambda -{\cal M}^2(m^2)\Big]h_{\mu\nu}=0,
\end{eqnarray}
where the mass squared ${\cal M}^2$ of a massive spin-2 graviton is
given by \be {\cal M}^2(m^2)=m^2-\frac{1}{2\ell^2} \to \frac{{\cal
M}^2}{m^2}=1-\frac{1}{2m^2\ell^2}. \ee Eq.(\ref{nmgmeq2}) describes
a massive graviton with 2$(6-4=2)$ DOF propagating around the BTZ
black hole under the gauge, while Eq.(\ref{nmgmeq1}) indicates a
non-propagating spin-2 graviton in the Einstein gravity. This
explains clearly why the NMG describes a massive graviton with 2
DOF. The presence of $S_{\rm FOT }$ distinguishes the NMG  from the
Einstein gravity because it generates 2 DOF.  At this stage, we
briefly mention the stability of the BTZ black hole in the NMG.  The
stability condition of the BTZ black hole in the NMG turned out to
be $m^2\ell^2>1/2({\cal M}^2>0)$ regardless of the horizon size
$r_+$, while the instability  condition is given by $ m^2\ell^2
<1/2({\cal M}^2<0)$~\cite{Moon:2013lea}. This is valid for the NMG,
not for the Einstein gravity.

Now we derive all thermodynamic quantities.  The Hawking temperature
is found to be  \be T_{\rm H}= \frac{f'(r_+)}{4\pi}=\frac{r_+}{2\pi
\ell^2} \ee which is the same for the Einstein gravity. Using the
ADT method, one could derive the mass~\cite{Clement:2009gq}, heat
capacity, entropy~\cite{Kim:2013qra}, and on-shell  free energy
\begin{eqnarray}  M_{\rm
ADT}=\frac{{\cal M}^2}{m^2}M,~~C_{\rm ADT}=\frac{{\cal
M}^2}{m^2}C,~~ S_{\rm ADT}=\frac{{\cal M}^2}{m^2}S_{\rm BH},~~F^{\rm
on}_{\rm ADT}=\frac{{\cal M}^2}{m^2}F^{\rm on}
\end{eqnarray} For $G=1/8$,  thermodynamic quantities in Einstein
gravity are given by~\cite{Myung:2005ee,Myung:2006sq,Eune:2013qs}
\be \label{btz} M=\frac{r_+^2}{\ell^2},~C=4\pi r_+,~S_{\rm BH}=
 4\pi r_+,~F^{on}= M-T_{\rm H} S_{\rm BH}=-\frac{r_+^2}{\ell^2}=-M\ee
which are positive regardless of the horizon size $r_+$ except that
the free energy is  negative. This means that the BTZ black hole is
thermodynamically stable in Einstein gravity.  Here we check that
the first-law of thermodynamics is satisfied as
\be\label{first-law}dM_{\rm ADT}=T_{\rm H} dS_{\rm ADT}, \ee as in
Einstein gravity
 \be
dM=T_{\rm H} dS_{\rm BH} \ee where `$d$' denotes the differentiation
with respect to the horizon size $r_+$ only.  Importantly, we note
that in the limit of $m^2 \to \infty$ we recovers thermodynamics of
the BTZ black hole in Einstein gravity, while in the limit of $m^2
\to 0$ we recover the black hole thermodynamics in purely
fourth-order gravity. The latter is similar to recovering the
third-order terms of  conformal Chern-Simons gravity  from the
topologically massive gravity~\cite{Bagchi:2013lma} and conformal
gravity from the Einstein-Weyl
gravity~\cite{Lu:2012xu,Myung:2013uka}.

It is well known that  the local thermodynamic stability is
determined by the positive heat capacity ($C_{\rm ADT}>0$) and the
global stability is determined by the negative free energy
($F^{on}_{\rm ADT}<0$). Therefore, we propose that the thermodynamic
stability is determined by the sign of the heat capacity while the
phase transition is mainly determined by the sign of the free
energy.

To investigate a phase transition, we introduce the thermal soliton
(TSOL) whose  thermodynamic quantities are given
by~\cite{Zhang:2015wna} \be \label{tads} M^{\rm TSOL}_{\rm
ADT}(m^2)=\frac{{\cal M}^2}{m^2}M^{\rm TSOL},~F^{\rm TSOL}_{\rm
ADT}(m^2)=\frac{{\cal M}^2}{m^2}F^{\rm TSOL},\ee where \be M^{\rm
TSOL}=1,~ F^{\rm TSOL}=-1 \ee with $G=1/8$. We note that a factor of
${\cal M}^2/m^2$ was missed in Ref.\cite{Myung:2013uka}. The TSOL
corresponds to the spacetime picture of the NS-NS vacuum
state~\cite{Maldacena:1998bw}.

Furthermore, we need the massless BTZ black hole (MBTZ) whose
thermodynamic quantities all are zero
as~\cite{Myung:2006sq,Rao:2007zzb} \be M^{\rm MBTZ}_{\rm
ADT}=0,~~F^{\rm MBTZ}_{\rm ADT}=0. \ee The MBTZ is called the
spacetime picture of the R-R vacuum state.

 In addition to
a global mass parameter $m^2$, we introduce five parameters
to describe  the phase transition in NMG. These are included as \ba &&\bullet~M:~{\rm order~ parameter}, \nonumber \\
&&\bullet~ T_{\rm H}(M) :~ {\rm order~ parameter~(onshell~temperature)}, \nonumber \\
&&\bullet~ T :~ {\rm control~ parameter~(offshell~temperature)}, \nonumber \\
&&\bullet~ F^{\rm on}_{\rm ADT}(m^2,M) :~ {\rm increasing~(decreasing)~black~hole~via~equilibrium~process}, \nonumber \\
&&\bullet~ F^{\rm off}_{\rm ADT}(m^2,M,T) :~{\rm
increasing~(decreasing)~black~hole~via~nonequilibrium~process},
\nonumber  \ea where off-shell (on-shell) mean equilibrium
(non-equilibrium) configurations. In general, the equilibrium
process implies a reversible process, while the non-equilibrium
process implies a irreversible process.  The off-shell free energy
corresponds to  a generalized free energy which is similar to a
temperature-dependent scalar potential $V(\varphi,T)$ for a simple
model of thermal phase transition where $\varphi$ is the order
parameter and $T$ is a control parameter.

\section{Phase transitions}
\subsection{HP phase transition}
\begin{figure}[t!]
   \centering
   \includegraphics{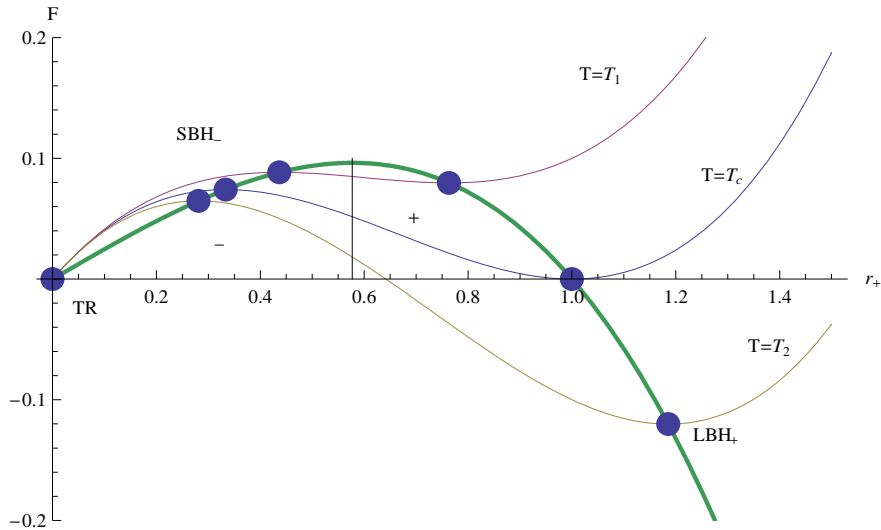}
\caption{Hawking-Page phase transition for the SAdS with $l=1$: the
thick curve represents the on-shell free energy $F^{\rm on}_{\rm
SAdS}(r_+)$, while three thin curves denote the off-shell free
energy $F_{\rm SAdS}^{\rm off}(r_+,T)$ with three temperatures
$T=T_1=0.9\pi^{-1}(<T_c),~T_c=\pi^{-1},T_2=1.1\pi^{-1}(>T_c)$.
$-(+)$ denote negative (positive) heat capacity.  The SBH$_-$ is
bounded in the right by the line at $r_+=r_*=0.57$. } \label{fig.7}
\end{figure}
Before we proceed, we understand  intuitively  how the original HP
transition occurs between Schwarzschild-AdS$_4$ black hole (SAdS)
and thermal radiation (TR).  For $G_4=1$, the ADM mass, Hawking
temperature, and the Bekenstein-Hawking entropy are given by
\begin{equation}
\label{aas2} M_{\rm
SAdS}(r_+)=\frac{1}{2}\Big(r_++\frac{r_+^3}{l^2}\Big),~~T_{\rm
H}(r_+) = \frac{1}{4\pi} \Big( \frac{1}{r_+} + \frac{3r_+}{l^2}
\Big),~~S_{\rm BH}=\pi r_+^2.
\end{equation}
In addition,  the heat capacity and on-shell free energy are given
by
\begin{eqnarray}\label{aaSc}
C_{\rm SAdS}(r_+)= 2\pi r_+^2 \Bigg(\frac{3r_+^2+l^2 }{3r_+^2 - l^2
}\Bigg),~~F^{\rm on}_{\rm SAdS}(r_+)&=& \frac{r_{+}}{4}\Big(1 -
\frac{r_{+}^2}{l^2}\Big),
\end{eqnarray}
where $C_{\rm SAdS}$ blows up at $r_+=r_*=l/\sqrt{3}$ (heat capacity
is changed from $-\infty$ to $\infty$ at $r_+=r_*$). The critical
temperature $T_c=T_H(r_+)|_{r_+=r_c}=\frac{1}{\pi l}$ is determined
from  the condition of $F^{on}_{\rm SAdS}(r_+)=0$ for $r_+=r_c=l$.
The TR  is located at $r_+=0$ in this picture.

In studying thermodynamic stability, two relevant quantities are the
heat capacity $C_{\rm SAdS}$ which determines thermally local
stability (instability) for $C_{\rm SAdS}>0(C_{\rm SAdS}<0)$ and
on-shell free energy $F_{\rm SAdS}^{\rm on}$ which determines  the
thermally global stability (instability) for $F_{\rm SAdS}^{\rm
on}<0(_{\rm SAdS}^{\rm on}>0)$. A SAdS  is thermodynamically stable
only if  $C_{\rm SAdS}>0$ and $F^{\rm on}_{\rm SAdS}<0$.
  For simplicity, we choose $l=1$.  We
observe that the on-shell free energy (thick curve in Fig. 1) is
maximum at $r_+=r_*=0.57$ and zero at $r_+=r_c=1$ which determines
the critical temperature. For $r_+>r_c$, one finds negative free
energy.

In order to investigate  the HP phase transition, one has to
introduce the off-shell free energy as a function of $r_+$ and $T$
\begin{equation}
F^{\rm off}_{\rm SAdS}(r_+,T)=M_{\rm SAdS}(r_+)-T S_{\rm BH}(r_+),
\end{equation}
where $T$ plays a role of control parameter for taking a phase
transition.

For $T=T_2>T_c$, the process of phase transition is shown in Fig. 1
explicitly.  In this case, one starts with TR ($\bullet$) at $r_+=0$
in AdS space and a small black hole ($\bullet$:SBH$_-$)  appears  at
$r_+=0.28$.   Here the SBH$_-$ denotes a unstable small black hole
with $C_{\rm SAdS}<0$ and $F^{on}_{\rm SAdS}>0$. This plays a role
of the mediator.  Then, since the heat capacity changes from
$-\infty$ to $\infty$ at $r_+=r_*$, the large black hole
($\bullet$:LBH$_+$) finally comes out as a stable object at
$r_+=1.19$. Here the LBH$_+$ represents  a globally stable black
hole because of $C_{\rm SAdS}>0$ and $F^{on}_{\rm SAdS}<0$.
Actually, there is a change of the dominance at the critical
temperature $T=T_c$: from TR  to SAdS~\cite{HP}. This is called the
HP phase transition as a typical example of the first-order
transition in the gravitational system: TR$\to$ SBH$_-$$\to$
LBH$_+$.

For $T=T_1(<T_c)$, the free energy $F^{on}_{\rm SAdS}(0)=0$ of TR is
the lowest  state, while for the $T=T_2(>T_c)$ case, the lowest
state is the free energy $F^{on}_{\rm SAdS}(1.19)<0$ for the large
black hole.  Hence, for $T_1<T_c$, the ground state is TR, whereas
for $T_2>T_c$, the ground state is  LBH$_-$. There is no phase
transition for $T=T_1$.

\subsection{HP transition in NMG}

The off-shell free energy for the BTZ black hole is given by \be
\label{off-fbtz} F^{\rm off}_{\rm ADT}(m^2,M,T)=\frac{{\cal
M}^2}{m^2} \Bigg(M- 4\pi \ell T \sqrt{ M}\Bigg). \ee On the other
hand, the off-shell free energy for TSOL takes the
form~\cite{Eune:2013qs,Zhang:2015wna} \be \label{off-fsol}F^{\rm
off}_{\rm TSOL}(m^2,M)=\frac{{\cal
M}^2}{m^2}\Big(-M-2\sqrt{-M}\Big). \ee It is worth noting that the
author has claimed that the phase transition from TSOL to BTZ is
possible to occur without introducing
(\ref{off-fsol})~\cite{Myung:2006sq}. However, the construction of
(\ref{off-fsol}) is necessary to show how the transition from TSOL
to BTZ occurs nicely~\cite{Eune:2013qs}.

 Here we would like to mention that for
${\cal M}^2>0$, the BTZ under consideration are thermally stable
because the  heat capacity is always  positive.   Hence, the free
energy plays a key role in studying a phase transition between two
gravitational configurations.
\begin{figure}[t!]
   \centering
   \includegraphics{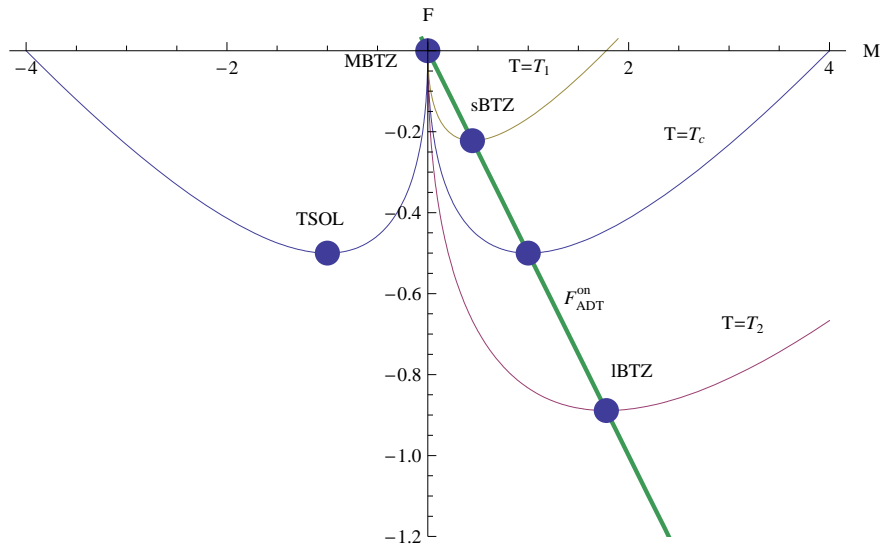}
\caption{On-shell free energy $F^{\rm on}_{\rm ADT}(m^2=1,M)=-M/2$
for the BTZ  and $F^{\rm TSOL}_{\rm ADT}(m^2=1)=-0.5$ for TSOL with
$G_3=1/8$ and $\ell=1$. Three off-shell free energies for the BTZ
are given by $F^{\rm off}_{\rm ADT}(m^2=1,M,T)$ for
$T=T_1=\frac{1}{3\pi},T_c=\frac{1}{2\pi},T_2=\frac{2}{3\pi}$. On the
other hand, the off-shell free energy for TSOL is $F^{\rm off}_{\rm
TSOL}$. They are connected at the point of  $M=0$. Here,  the MBTZ
located at $M=0$ plays a role of the mediator like SBH$_-$. }
\end{figure}
We emphasize  that two on-shell free energies $F^{on}_{\rm ADT}$ and
$F^{\rm TSOL}=-1/2$ are disconnected to each other.   To discuss the
phase transition with ${\cal M}^2>0(m^2=1>1/2)$, we would be better
to  examine two off-shell free energies (\ref{off-fbtz}) and
(\ref{off-fsol}).  One finds from Fig. 2 that for $T=T_1$, the TSOL
($\bullet$) is more favorable than the small BTZ ($\bullet$), while
for $T=T_2$, the large BTZ ($\bullet$) is more favorable than TSOL
($\bullet$). This observation suggests a phase transition
(TSOL$\to$MBTZ$\to$BTZ) for $T>T_c$.  Two off-shell free energies
are connected at the point $M=0$. At $T=T_c$, the transition from
TSOL to BTZ black hole through MBTZ is possible to occur. For
$T=T_1<T_c$, the TSOL dominates because of $F^{\rm TSOL}<F^{on}_{\rm
ADT}$, while for $T=T_2>T_c$, the BTZ dominates because of $F^{\rm
TSOL}>F^{on}_{\rm ADT}$. This indicates that a change of dominance
occurs at the critical temperature $T=T_c$~\cite{Myung:2006sq}.
Importantly, this transition  could be regarded really  as a HP
transition because the MBTZ plays a role of the mediator like
SBH$_-$ in the previous  HP phase transition~\cite{Myung:2013uka}. A
difference is that the MBTZ is a single extremal state (see Fig.2),
whereas SBH$_-$ has three states depending on $T$ (see Fig.1). A
similarity is that they all  are the highest states.

\subsection{Inverse HP transition in NMG}
Generally, increasing black hole ($\searrow$) is induced  by
absorbing radiations in the heat reservoir, while decreasing black
hole ($\swarrow$) is done by Hawking radiations as evaporation
process. The former describes the HP phase transition, whereas the
latter denotes the inverse Hawking-Page phase transition (IHP).
\begin{figure}[t!]
   \centering
   \includegraphics{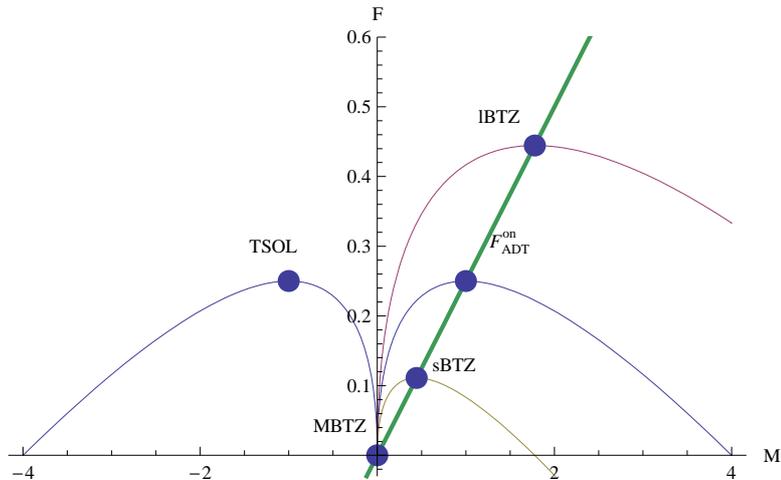}
\caption{On-shell free energy $F^{\rm on}_{\rm ADT}(m^2=0.4,M)=0.25
M$ for the BTZ  and $F^{\rm TSOL}_{\rm ADT}(m^2=0.4)=0.25$ for TSOL.
Three off-shell free energies for the BTZ are given by $F^{off}_{\rm
ADT}(m^2=0.4,M,T)$ for
$T=T_1=\frac{1}{3\pi},T_c=\frac{1}{2\pi},T_2=\frac{2}{3\pi}$. On the
other hand, the off-shell free energy for TSOL is $F^{\rm off}_{\rm
TSOL}(m^2=0.4,M)$. Here we observe that the MBTZ located at $M=0$ is
the ground state.}
\end{figure}

For ${\cal M}^2<0$ case,  fourth-order curvature terms (\ref{NFO})
contribute dominantly to black hole thermodynamics. In this case,
the heat capacity of the BTZ is negative and the on-shell free
energy is positive, which means that the BTZ is thermodynamically
unstable. This is consistent with the classical instability of the
BTZ for $m^2<1/2$~\cite{Moon:2013lea}.   The other quantities of ADT
mass and entropy are negative, which may raise a problem in
obtaining a consistent black hole thermodynamics. To avoid this
problem, the author in~\cite{Zhang:2015wna} has required ${\cal
M}^2>0$. However, at this moment, we could not understand why this
problem arises in the ADT approach to the black hole thermodynamics
in the NMG. Instead, we would be better to  make a progress on the
phase transition because this is a main feature of the BTZ in the
NMG. Here, we wish to know whether a phase transition from the BTZ
to TSOL is possible to occur in NMG for ${\cal M}^2<0(m^2=0.4<1/2)$.
We find from Fig. 3 that the MBTZ located at $M=0$ is always more
favorable than BTZ and TSOL because of $F_{\rm MBTZ}<F_{\rm
ADT}^{\rm on}(m^2=0.4,M),F^{\rm TSOL}(m^2=0.4)$. In other words, it
corresponds to  the ground state.  In this case, we might not define
a possible phase transition between TSOL and BTZ because the ground
state is  given by the MBTZ. A possible transition is the IHP
(BTZ$\to$MBTZ)~\cite{Myung:2006sq}, whereas  the TSOL is located at
the left branch. This may be possible because the MBTZ is considered
as an  extremal black hole without size.

\section{Discussions}
We have investigated  thermodynamics of the BTZ black hole in new
massive gravity completely. We have confirmed that for
$m^2\ell^2>1/2$, the Hawking-Page phase transition occurs between
the BTZ black hole and thermal soliton by introducing the massless
BTZ black hole. Here the massless BTZ black hole plays a role of the
mediator like SBH$_-$ in the original Hawking-Page phase transition
and it is the highest state.

  On the
other hand, for $m^2\ell^2<1/2$, this transition unlikely occurs but
a phase transition between the BTZ black hole and the massless BTZ
black hole is possible to occur. In this case, the massless BTZ
black hole is the ground state.  We call the latter as the inverse
Hawking-Page phase transition and this transition is favored in the
new massive gravity. This completes phase transitions of the BTZ
black hole in the new massive gravity.
\section*{Acknowledgement}

 This work was supported by the 2015 Inje University  research grant.


\begin{thebibliography}{99}

\bibitem{HP} S. W. Hawking  and  D. N. Page, Commun. Math. Phys. {\bf
87} (1983) 577.
%%CITATION = CMPHA,87,577;%%

\bibitem{BCM} J.~D.~Brown, J.~Creighton and R.~B.~Mann,
  %``Temperature, energy and heat capacity of asymptotically anti-de Sitter
  %black holes,''
  Phys.\ Rev.\ D {\bf 50} (1994) 6394
  [gr-qc/9405007].
  %%CITATION = GR-QC 9405007;%%

  \bibitem{Witt}
  E.~Witten,
  %``Anti-de Sitter space, thermal phase transition, and confinement in  gauge
  %theories,''
  Adv.\ Theor.\ Math.\ Phys.\  {\bf 2} (1998) 505
  [hep-th/9803131].
  %%CITATION = HEP-TH 9803131;%%

%\cite{Arnowitt:1962hi}
\bibitem{Arnowitt:1962hi}
  R.~L.~Arnowitt, S.~Deser and C.~W.~Misner,
  %``The Dynamics of general relativity,''
  Gen.\ Rel.\ Grav.\  {\bf 40}, 1997 (2008)  [gr-qc/0405109].
  %%CITATION = GR-QC/0405109;%%
  %585 citations counted in INSPIRE as of 31 Oct 2013


\bibitem{dbranes}  C. V. Johnson, {\it D-Branes}, Cambbridge Univ.
Press (2003, Camambridge).

%\cite{Deser:1981wh}
\bibitem{Deser:1981wh}
  S.~Deser, R.~Jackiw and S.~Templeton,
  %``Topologically Massive Gauge Theories,''
  Annals Phys.\  {\bf 140}, 372 (1982)  [Annals Phys.\  {\bf 185}, 406 (1988)]  [Annals Phys.\  {\bf 281}, 409 (2000)].
  %%CITATION = APNYA,140,372;%%  %2286 citations counted in INSPIRE as of 12 Jun 2015

%\cite{Bergshoeff:2009hq}
\bibitem{Bergshoeff:2009hq}
  E.~A.~Bergshoeff, O.~Hohm and P.~K.~Townsend,
  %``Massive Gravity in Three Dimensions,''
  Phys.\ Rev.\ Lett.\  {\bf 102}, 201301 (2009)  [arXiv:0901.1766 [hep-th]].
  %%CITATION = ARXIV:0901.1766;%%  %383 citations counted in INSPIRE as of 12 juin 2015


%\cite{Abbott:1981ff}
\bibitem{Abbott:1981ff}
  L.~F.~Abbott and S.~Deser,
  %``Stability of Gravity with a Cosmological Constant,''
  Nucl.\ Phys.\ B {\bf 195}, 76 (1982).
   %%CITATION = NUPHA,B195,76;%%
    %555 citations counted in INSPIRE as of 31 Oct 2013

%\cite{Deser:2002jk}
\bibitem{Deser:2002jk}
  S.~Deser and B.~Tekin,
  %``Energy in generic higher curvature gravity theories,''
  Phys.\ Rev.\ D {\bf 67}, 084009 (2003)  [hep-th/0212292].
  %%CITATION = HEP-TH/0212292;%%
   %181 citations counted in INSPIRE as of 31 Oct 2013

%\cite{Kim:2013zha}
\bibitem{Kim:2013zha}
  W.~Kim, S.~Kulkarni and S.~-H.~Yi,
  %``Quasi-Local Conserved Charges in Covariant Theory of Gravity,''
   Phys.\ Rev.\ Lett.\  {\bf 111}, 081101 (2013)  [arXiv:1306.2138 [hep-th]].
    %%CITATION = ARXIV:1306.2138;%%  %2 citations counted in INSPIRE as of 31 Oct 2013


%\cite{Zhang:2015wna}
\bibitem{Zhang:2015wna}
  S.~J.~Zhang,
  %``Hawking?Page phase transition in new massive gravity,''
   Phys.\ Lett.\ B {\bf 747}, 158 (2015).  %%CITATION = PHLTA,B747,158;%%

%\cite{Moon:2013lea}
\bibitem{Moon:2013lea}
  T.~Moon and Y.~S.~Myung,
  %``Gregory-Laflamme instability of the BTZ black hole in new massive gravity,''
   Phys.\ Rev.\ D {\bf 88}, no. 12, 124014 (2013)  [arXiv:1310.3024 [hep-th]].
    %%CITATION = ARXIV:1310.3024;%%  %1 citations counted in INSPIRE as of 10 juin 2015

%\cite{Myung:2013uka}
\bibitem{Myung:2013uka}
  Y.~S.~Myung and T.~Moon,
  %``Thermodynamic and classical instability of AdS black holes in fourth-order gravity,''
  JHEP {\bf 1404}, 058 (2014)  [arXiv:1311.6985 [hep-th]].
  %%CITATION = ARXIV:1311.6985;%%  %1 citations counted in INSPIRE as of 10 juin 2015



\bibitem{BTZ-1}  M.~Banados, C.~Teitelboim and J.~Zanelli,
  %``The Black hole in three-dimensional space-time,''
  Phys.\ Rev.\ Lett.\  {\bf 69}, 1849 (1992)
  [arXiv:hep-th/9204099].
  %%CITATION = PRLTA,69,1849;%%

\bibitem{BTZ-2} M.~Banados, M.~Henneaux, C.~Teitelboim and J.~Zanelli,
  %``Geometry of the (2+1) black hole,''
  Phys.\ Rev.\  D {\bf 48}, 1506 (1993)
  [arXiv:gr-qc/9302012].
  %%CITATION = PHRVA,D48,1506;%%

%\cite{Clement:2009gq}
\bibitem{Clement:2009gq}
  G.~Clement,
  %``Warped AdS(3) black holes in new massive gravity,''
  Class.\ Quant.\ Grav.\  {\bf 26}, 105015 (2009)  [arXiv:0902.4634 [hep-th]].
  %%CITATION = ARXIV:0902.4634;%%  %75 citations counted in INSPIRE as of 31 Oct 2013


%\cite{Kim:2013qra}
\bibitem{Kim:2013qra}
  W.~Kim, S.~Kulkarni and S.~-H.~Yi,
  %``Conserved quantities and Virasoro algebra in New massive gravity,''
   JHEP {\bf 1305}, 041 (2013)  [arXiv:1303.3691 [hep-th]].
    %%CITATION = ARXIV:1303.3691;%%  %4 citations counted in INSPIRE as of 31 Oct 2013




%\cite{Myung:2005ee}
\bibitem{Myung:2005ee}
  Y.~S.~Myung,
  %``No Hawking-Page phase transition in three dimensions,''
  Phys.\ Lett.\ B {\bf 624}, 297 (2005)  [hep-th/0506096].
  %%CITATION = HEP-TH/0506096;%%  %18 citations counted in INSPIRE as of 31 Oct 2013



%\cite{Myung:2006sq}
\bibitem{Myung:2006sq}
  Y.~S.~Myung,
  %``Phase transition between the BTZ black hole and AdS space,''
  Phys.\ Lett.\ B {\bf 638}, 515 (2006)  [gr-qc/0603051].
  %%CITATION = GR-QC/0603051;%%  %16 citations counted in INSPIRE as of 31 Oct 2013

%\cite{Eune:2013qs}
\bibitem{Eune:2013qs}
  M.~Eune, W.~Kim and S.~-H.~Yi,
  %``Hawking-Page phase transition in BTZ black hole revisited,''
  JHEP {\bf 1303}, 020 (2013)  [arXiv:1301.0395 [gr-qc]].
  %%CITATION = ARXIV:1301.0395;%%  %3 citations counted in INSPIRE as of 31 Oct 2013

%\cite{Bagchi:2013lma}
\bibitem{Bagchi:2013lma}
  A.~Bagchi, S.~Detournay, D.~Grumiller and J.~Simon,
  %``Cosmic evolution from phase transition of 3-dimensional flat space,''
  Phys.\ Rev.\ Lett.\  {\bf 111}, 181301 (2013)  [arXiv:1305.2919 [hep-th]].
   %%CITATION = ARXIV:1305.2919;%%  %1 citations counted in INSPIRE as of 31 Oct 2013



%\cite{Lu:2012xu}
\bibitem{Lu:2012xu}
  H.~Lu, Y.~Pang, C.~N.~Pope and J.~F.~Vazquez-Poritz,
  %``AdS and Lifshitz Black Holes in Conformal and Einstein-Weyl Gravities,''
   Phys.\ Rev.\ D {\bf 86}, 044011 (2012)  [arXiv:1204.1062 [hep-th]].
   %%CITATION = ARXIV:1204.1062;%%  %29 citations counted in INSPIRE as of 31 Oct 2013

%\cite{Maldacena:1998bw}
\bibitem{Maldacena:1998bw}
  J.~M.~Maldacena and A.~Strominger,
  %``AdS(3) black holes and a stringy exclusion principle,''
  JHEP {\bf 9812}, 005 (1998)  [hep-th/9804085].
   %%CITATION = HEP-TH/9804085;%%  %537 citations counted in INSPIRE as of 12 juin 2015

%\cite{Rao:2007zzb}
\bibitem{Rao:2007zzb}
  X.~P.~Rao, B.~Wang and G.~H.~Yang,
  %``Quasinormal modes and phase transition of black holes,''
  Phys.\ Lett.\ B {\bf 649}, 472 (2007)  [arXiv:0712.0645 [gr-qc]].
  %%CITATION = ARXIV:0712.0645;%%  %18 citations counted in INSPIRE as of 12 juin 2015


\end{thebibliography}
\end{document}